# Lithium niobate micro-disk resonators of quality factors above 10$^7$

RONGBO WU,[1,4,†] JIANHAO ZHANG,[1,4,†] NI YAO,[3] WEI FANG,[3] LINGLING QIAO,[1] ZHIFANG CHAI,[2] JINTIAN LIN,[1,7] AND YA CHENG[1,2,4,5,6,*]

[1] *State Key Laboratory of High Field Laser Physics, Shanghai Institute of Optics and Fine Mechanics, Chinese Academy of Sciences, Shanghai 201800, China*
[2] *State Key Laboratory of Precision Spectroscopy, East China Normal University, Shanghai 200062, China*
[3] *State Key Laboratory of Modern Optical Instrumentation, College of Optical Science and Engineering, Zhejiang University, Hangzhou 310027, China*
[4] *University of Chinese Academy of Sciences, Beijing 100049, China*
[5] *XXL-The Extreme Optoelectromechanics Laboratory, School of Physics and Materials Science, East China Normal University, Shanghai 200241, China*
[5] *Collaborative Innovation Center of Extreme Optics, Shanxi University, Taiyuan, Shanxi 030006, China*
[7] e-mail: jintianlin@siom.ac.cn
*Corresponding author: ya.cheng@siom.ac.cn

**We report on fabrication of crystalline lithium niobate microresonators with quality factors above 10$^7$ as measured around 770 nm wavelength. Our technique relies on femtosecond laser micromachining for patterning a mask coated on the lithium niobate on insulate (LNOI) into a microdisk, followed by a chemo-mechanical polishing process for transferring the disk-shaped pattern to the LNOI. Nonlinear processes including second harmonic generation and Raman scattering have been demonstrated in the fabricated microdisk.**

The realization of micro-disk resonators (MDRs) of high quality (Q) factors using lithium niobate on insulator (LNOI) as the substrate has spurred great interest in developing on-chip nanophotonic structures which hold the promise for efficient nonlinear wavelength conversion, fast electrooptic light modulation, and high density photonic integration [1-3]. The key to achieve the high Q factors in the fabricated lithium niobate (LN) MDRs is the incorporation of focused ion beam (FIB) milling or ion dry etching in the patterning of the LNOI, which gives rise to smooth sidewalls for minimizing the scattering loss [4-18]. Typical Q factors on the level of 10$^6$ can now be routinely obtained for freestanding LN MDRs [7-10]. Moreover, it is recently reported that a Q factor of 3×10$^6$ measured around 976 nm wavelength can be achieved for the LN MDRs fabricated by a reactive ion etching followed by a chemo-mechanical (CM) polishing process, and the intrinsic Q factor could have possibly reached 10$^7$ around the optical communication wavelength of 1.5 μm [10]. The CM polishing greatly improves the surface smoothness, leading to a significant increase of the Q factor.

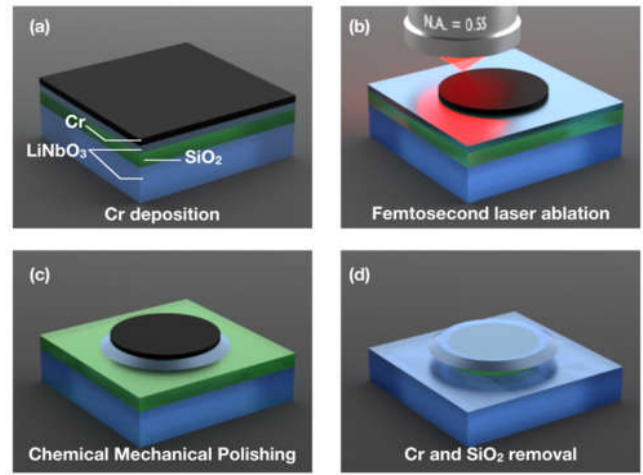

Fig. 1. Illustration of fabrication flow. (a) Coating a Cr thin film on top of the LNOI. (b) Patterning the Cr thin film into a microdisk. (c) Transferring the disk-shaped pattern to LNOI by chemo-mechanical polishing. (d) Removing the Cr thin film and the SiO$_2$ buffer layer with two consecutive chemical wet etching processes.

Here, we show that by combining the femtosecond laser direct writing and the CM polishing, we are able to realize LN MDRs of Q factors exceeding $10^7$ near the wavelength of 773 nm. This is almost one order of magnitude higher than the state-of-the-art Q factors around the visible and near-infrared wavelengths reported so far [7-10]. More importantly, since we now can completely skip the FIB milling process, the fabrication throughput can be dramatically boosted, and fabrication of large scale photonic chips beyond the range of motion of the FIB is readily achievable. Although we do not investigate the combination of other mass-fabrication technologies based on the ion dry etching or ultraviolet lithography with the CM polishing, similar Q factors are expected for the LN MDRs obtained with such approaches. The high Q LN MDRs fabricated with the fast and flexible laser writing technology will provide a fascinating platform for LNOI photonics application.

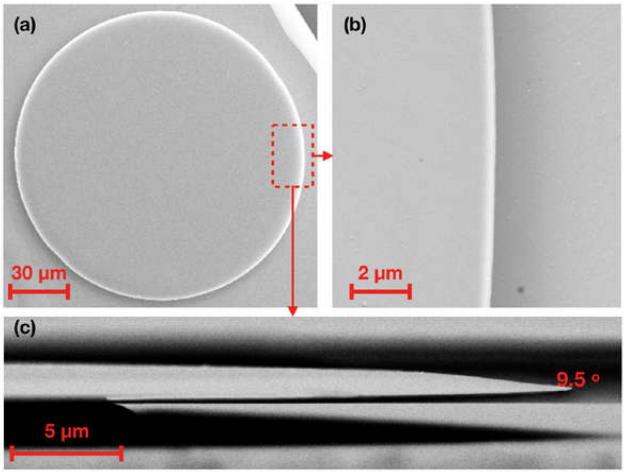

Fig. 2. (a) Top view SEM image of a fabricated LN MDR. (b) Close up view SEM image of the area indicated by the red box in (a). (c) Side view SEM image of the fabricated LN MDR with a wedge angle of 9.5°.

In our experiment, the LN MDRs were produced on a commercially available X-cut LN thin film wafer fabricated by ion slicing (NANOLN, Jinan Jingzheng Electronics Co., Ltd) [1]. The LN thin film with a thickness of 900 nm is bonded to a 2 μm thick $SiO_2$ layer grown on a LN substrate. The fabrication process includes four steps, as schematically illustrated in Fig. 1. First, a thin layer of chromium (Cr) with a thickness of 900 nm was deposited on the surface of the LNOI by thermal evaporation coating. Subsequently, the Cr film on the LNOI sample was patterned into a circular disk using space-selective femtosecond laser direct writing. Specifically, to minimize the heat effect as well as the redeposition of the ablation debris on the disk surface, the femtosecond laser ablation was conducted by immersing the LNOI sample coated with Cr in water. The femtosecond laser pulses were focused at the Cr surface with an objective lens (NA= 0.55, Nikon LU Plan). The key in this step is to carefully choose a pulse energy of the femtosecond laser so as to enable a complete removal of the Cr film with laser ablation while keeping the underneath LNOI intact thanks to the high precision and low heat generation in the interaction of femtosecond laser pulses with materials [19]. The Cr disk formed on the LNOI serves as a hard mask for the subsequent CM polishing. Next, the CM polishing process was performed to fabricate the LN MDRs with a wafer polishing machine (NUIPOL802, Kejing, Inc.). In this case, the top surface of the LNOI defined by the femtosecond laser patterning was protected from the CM polishing because of the Cr hard mask, whereas as the open area of the LNOI without the protection from the Cr film can be accessed and in turn completed removed by the polishing slurry (MasterMet, Buehler, Ltd.). Therefore, the disk-shaped pattern was transferred to the LNOI with a smooth sidewall. Finally, the fabricated structure was first immersed in a Cr etching solution (Chromium etchant, Alfa Aesar GmbH) for 10 min, and then underwent a chemical wet etching in a buffered hydrofluoric acid (HF) solution (BUFFER HF IMPROVED, Transene Co., Inc.) to partially remove the $SiO_2$ layer beneath the LN microdisk. The LN microdisk supported by the $SiO_2$ pedestal was produced to form the freestanding LN MDRs. Figure 2 shows the scanning electron micrograph (SEM) images of a fabricated LN MDR with a diameter of 140 μm.

To characterize the optical mode structure of the LN MDR, a tunable laser was used to couple light into and out of the fabricated MDR through a tapered fiber with a waist of 0.9 μm. The linewidth of the tunable laser (TLB 6712, New Focus, Inc.) is 200 kHz. The optical modes could be excited by controlling the relative position between the tapered fiber and the MDR. The transmission power of the tapered fiber coupled with the MDR was record by a transient optic receiver (1801-FC, New Focus, Inc.). Figure 3(a) shows the transmission spectrum for the wavelength range from 772.8 to 773.9 nm. The free spectral range (FSR) of the MDR is determined to be 0.45 nm. One of the whispering-gallery modes at the resonant wavelength of 773.49 nm was chosen for the measurement of the Q factor by fitting with a Lorentz function, which reaches $1.46\times10^7$ as indicated by the Lorentz fitting curve in Fig. 3(b).

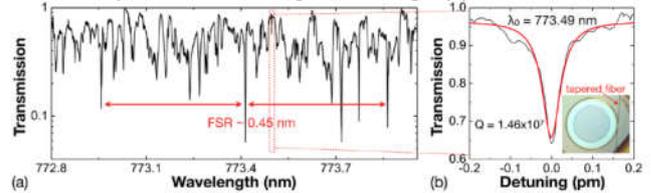

Fig. 3. (a) Transmission spectrum of the LN DMRs. (b) The Lorentz fitting (red curve) reveals a Q factor of $1.46\times10^7$. Inset: optical micrograph of the MDR coupling with the fiber taper.

The nonlinear optical properties of the high Q MDR was examined with another tunable laser, which was boosted by an erbium-ytterbium-doped fiber amplifier to serve as the pump source (EYDFA, Golight, Inc.). The tunable laser has a linewidth of 10 MHz and a wavelength tuning range between 1510 nm and 1620 nm. The pump laser power was adjusted by a variable optical attenuator. The polarization of the pump laser was controlled to have the quasi-TM (transverse-magnetic) polarization using an online fiber polarization controller. The emission signal from the MDR was record by a spectrometer (SR 303i, Andor, Inc.). Two pieces of short-pass filters (FESH1000, Thorlabs, Inc.) were used to block the pump laser during the spectral measurements of the nonlinear optical signals.

When we set the pump laser wavelength at 1561.2 nm and the pump laser power at 27 mW, cascaded nonlinear optical processes including the second harmonic generation (SHG) and Raman scattering mediated by the SHG were observed, as shown in Fig. 4(a). The SHG signal was detected to have a quasi-TE (transverse-electric) polarization at the wavelength of 780.6 nm. The vibrational modes participating the Raman scatterings were 152 $cm^{-1}$ and 239 $cm^{-1}$ [21], as indicated by the Raman peaks S1, and S2 in the vicinity of the SHG spectrum in Fig. 4(a). The conversion efficiency of the SHG as a function of the pump power is shown in Fig. 4(b). The normalize conversion efficiency of SHG can be determined as $2.3\times10^{-5}$/mW, which is lower than the previous result obtained in an X-cut MRD with a Q factor of $10^5$ [20]. Although the same cyclic phase matching scheme is employed in both the current and previous experiments of SHG, the relatively lower conversion efficiency is probably due to the small spatial overlapping between the pump and signal waves in the CM-polished LN MDRs with an extended disk wedge. Further theoretical and experimental investigations are required to reach a better understanding on the details in the nonlinear optical processes.

Furthermore, the powers of the Raman signals (i.e., S1 and S2) as a function of the SHG power are plotted in Fig. 4(c) and (d), respectively. The threshold pump powers (i.e., the power of SHG) for generating the Raman peaks S1 and S2 were determined to be 0.85 µW and 1.33 µW, respectively. Above the threshold powers of SHG, the powers of the Raman signals at S1 and S2 increases linearly with the pump power. From the slope of the fitting lines in Fig. 4(c) and (d), the conversion efficiencies as high as 0.28% and 0.22% have been determined. The conversion efficiencies are much higher than the SHG process, which agree well with our expectation for the reasons as follows. First, the Raman process does not require phase-matching, which is one of the major difficulties to overcome in achieving the high conversion efficiencies of the nonlinear processes in the whispering gallery mode microresonators. Second, the wavelengths of the two Raman peaks S1 and S2 are both close to the second harmonic wavelength, indicating that a sufficient spatial overlap can be realized for the modes of the SHG pump wave and the two Raman signal waves. All of these are beneficial for achieving the high conversion efficiencies.

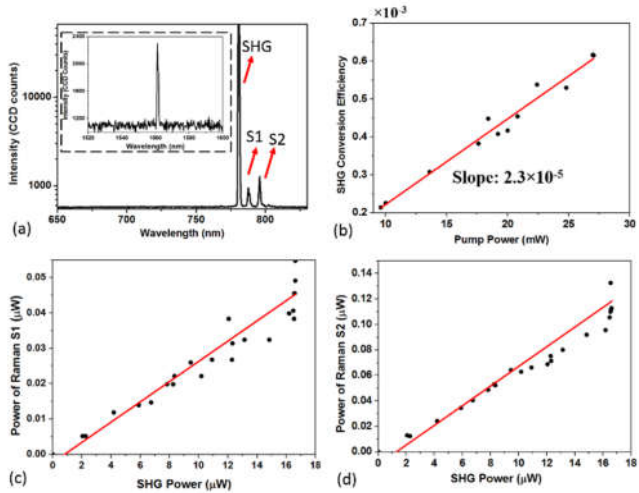

Fig. 4. (a) The nonlinear optical spectrum around the wavelength of 775 nm, Inset: the transmission spectrum around the pump wavelength. (b) The SHG conversion efficiency as a function of pump power. (c) The power of Raman S1 as a function of the SHG power. (d) The power of Raman S2 as a function of the SHG Power.

To conclude, on-chip LN MDRs of quality factors comparable to that of the mechanically polished crystalline optical resonators (i.e., $10^7$) have been achieved with the femtosecond laser micromachining combined with the CM polishing process. We stress that the femtosecond laser micromachining is only employed for patterning the protecting Cr thin film but not the LNOI. Essentially, the LN MDRs are created by transferring the patterns of the Cr layer to the LNOI in the CM polishing, leading to the ultrahigh Q factors even at the near-infrared wavelengths around 770 nm. This fabrication strategy should be effective by replacing the femtosecond laser direct writing in the patterning of the Cr film with other mass-scale lithography-based technologies such as ultraviolet lithography, electron beam lithography, *etc*.

**Funding.** National Basic Research Program of China (Grant No. 2014CB921303), National Natural Science Foundation of China (Grant Nos. 11734009, 61590934, 61635009, 61327902, 61505231, 11604351, 11674340, 61575211, 61675220), the Strategic Priority Research Program of Chinese Academy of Sciences (Grant No. XDB16000000), Key Research Program of Frontier Sciences, Chinese Academy of Sciences (Grant No. QYZDJ-SSW-SLH010), the Project of Shanghai Committee of Science and Technology (Grant 17JC1400400), Shanghai Rising-Star Program (Grant No. 17QA1404600), and the Fundamental Research Funds for the Central Universities.

†These authors contributed equally to this Letter.

**References**

1. G. Poberaj, H. Hu, W. Sohler, and P. Günter, Laser Photon. Rev. **6**, 488 (2012).
2. M. Wang, J. Lin, Y. Xu, Z. Fang, L. Qiao, Z. Liu, W. Fang, and Y. Cheng, Opt. Commun. **395**, 249 (2017).
3. A. Boes, B. Corcoran, L. Chang, J. Bowers, A. Mitchell, Laser Photon. Rev. **12**, 1700256 (2018).
4. J. Lin, Y. Xu, Z. Fang, J. Song, N. Wang, L. Qiao, W. Fang, and Y. Cheng, arXiv:1405.6473 (2014).
5. J. Lin, Y. Xu, Z. Fang, M. Wang, J. Song, N. Wang, L. Qiao, W. Fang, and Y. Cheng, Sci. Rep. **5**, 8072 (2015).
6. C. Wang, M. J. Burek, Z. Lin, H. A. Atikian, V. Venkataraman, I-C. Huang, P. Stark, and M. Lončar, Opt. Express **22**, 30924 (2014).
7. J. Wang, F. Bo, S. Wan, W. Li, F. Gao, J. Li, G. Zhang, and J. Xu, Opt. Express **23**, 23072 (2015).
8. J. Lin, Y. Xu, Z. Fang, M. Wang, J. Song, N. Wang, L. Qiao, W. Fang, and Y. Cheng, Sci. China. Mech. Astron. **58**, 14209 (2015).
9. R. Luo, H. Jiang, S. Rogers, H. Liang, Y. He, and Q. Lin, Opt. Express **25**, 24531 (2017).
10. R. Wolf, I. Breunig, H. Zappe, and K. Buse, Opt. Express **25**, 29927 (2017).
11. W. C. Jiang and Q. Lin, Sci. Rep. **6**, 36920 (2016).
12. S. Diziain, R. Geiss, M. Steinert, C. Schmidt, W. Chang, S. Fasold, D. Füßel, Y. Chen, and T. Pertsch, Opt. Mater. Express **5**, 2081 (2015).
13. M. Wang, Y. Xu, Z. Fang, Y. Liao, P. Wang, W. Chu, L. Qiao, J. Lin, W. Fang, and Y. Cheng. Opt. Express **25**, 124 (2017).
14. S. Liu, Y. Zheng, and X. Chen, Opt. Lett. **42**, 3626 (2017).
15. H. Jiang, R. Luo, H. Liang, X. Chen, Y. Chen, and Q. Lin, Opt. Lett. **42**, 3267 (2017).
16. Z. Hao, J. Wang, S. Ma, W. Mao, F. Bo, F. Gao, G. Zhang, and J. Xu, Photon. Res. **6**, 623 (2017).
17. J. Moore, J. K. Douglas, I. W. Frank, T. A. Friedmann, R. Camacho, and M. Eichenfield, Proc. Conf. Lasers and Electro-Optics (CLEO), pp. STh3P.1 (2016).
18. M. Wang, N. Yao, R. Wu, Z. Fang, S. Lv, J. Zhang, L. Qiao, J. Lin, W. Fang, Y. Cheng, arXiv:1803.11341 (2018).
19. K. Sugioka and Y. Cheng, Light Sci. Appl. **3**, e149 (2014).
20. J. Lin, Y. Xu, J. Ni, M. Wang, Z. Fang, L. Qiao, W. Fang, and Y. Cheng, Phys. Rev. Appl. **6**, 014002 (2016).
21. R. F. Schaufele and M. J. Weber, Phys. Rev. **152**, 705 (1966).